# Dream 2 Image

## An Open Multimodal EEG Dataset

### for Decoding and Visualizing Dreams

#### with AI

Yann Bellec

*Department of Neuroscience, University of California San Diego, 9500 Gilman Dr, La Jolla, CA 92093.*

**ABSTRACT**

Dream2Image is the world's first dataset combining EEG signals, dream transcriptions, and AI-generated images. Based on 38 participants and more than 31 hours of dream EEG recordings, it contains 129 samples offering: the final seconds of brain activity preceding awakening (T-15, T-30, T-60, T-120), raw reports of dream experiences, and a approximate visual reconstruction of the dream. This dataset provides a novel resource for dream research, a unique resource to study the neural correlates of dreaming, to develop models for decoding dreams from brain activity, and to explore new approaches in neuroscience, psychology, and artificial intelligence. Available in open access on Hugging Face and GitHub, Dream2Image provides a multimodal resource designed to support research at the interface of artificial intelligence and neuroscience. It was designed to inspire researchers and extend the current approaches to brain activity decoding. Limitations include the relatively small sample size and the variability of dream recall, which may affect generalizability. The Dream2Image dataset is openly available on Hugging Face at: https://huggingface.co/datasets/opsecsystems/Dream2Image

## 1 Introduction

Since the dawn of humanity, dreams have fascinated and puzzled the human mind. Yet, despite considerable advances in psychology and neuroscience, their study remains limited by the lack of multimodal data. With the explosive progress of artificial intelligence in recent years, research in brain decoding has shown that it is possible to generate images from brain activity using neuroimaging techniques such as EEG and fMRI. These approaches provide new opportunities to investigate altered states of consciousness, in particular REM sleep, during which most vivid dreaming typically occurs. Today, dream research suffers from a shortage of usable data. While EEG recordings of sleeping subjects and dream reports exist, they remain scarce and fragmented. Consequently, the scientific literature on dreams is fragmented and lacks a multimodal foundation.

To address this gap, we introduce Dream2Image, the first dataset integrating EEG signals recorded during sleep, faithful dream reports, and AI-generated images derived from these transcriptions. The dataset is based on a novel methodological approach in neuroscience. Its central aim is to establish a link between neuroscience, artificial intelligence, and psychology, particularly through brain imaging.

By making this dataset openly accessible, we provide the scientific community with a tool to explore new horizons: understanding the neural correlates of dreams, studying the relationship between brain activity and subjective experience, and developing AI models capable of visualizing the unconscious. Dream2Image provides a first attempt at systematically uniting neuroscience and artificial intelligence in the context of dream research. Similar research protocols have already been conducted using fMRI and the collection of experience reports in altered states of consciousness in the experiment by Horikawa et al. (2013). It is now established that dream reports reflect genuine conscious experiences during sleep, rather than confabulations upon awakening (Siclari et al., 2017).

## 2 Methodology

### 2.1 Participants

The EEG data were sourced from the **DREAM DATABASE** (Monash University, 2025), an international repository dedicated to sleep and dream research. Two main datasets were selected: (1) the work of Zhang & Wamsley (2019), which investigated EEG predictors of dreaming outside of REM sleep, and (2) the **Two-way Communication USA** dataset (Konkoly et al., 2021), which provided real-time communication data between experimenters and dreamers during REM sleep.

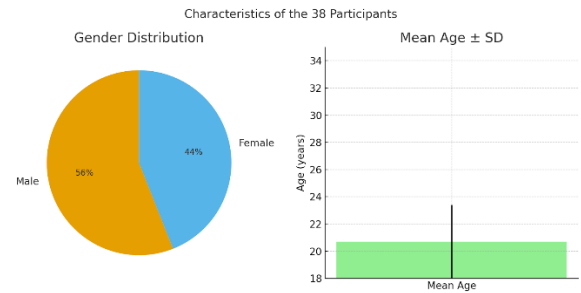

*Fig. 1 – Demographic and recruitment characteristics of the 38 participants included in the study (ages 18–35, mean 20.7 ± 2.7 years; 56% male, 44% female). Participants were recruited at Northwestern University and Furman University (USA) and met strict inclusion criteria (regular sleep schedule, absence of sleep/psychiatric disorders, no interfering medication).*

A total of 38 participants were included (ages 18–35, mean 20.7 ± 2.7 years; 56% male, 44% female). Participants were recruited via word-of-mouth and online calls from Northwestern University and Furman University (Greenville, USA).

Exclusion criteria included a history of sleep or psychiatric disorders, current medication interfering with sleep, and extensive prior experience with 3D video games (related to the objectives of the broader study).

Participants were also required to maintain a regular sleep schedule for three consecutive nights before the experiment, refrain from recreational drugs or alcohol 24 hours before the session, and avoid caffeine after 10 a.m. on the day of the study.

### 2.2 EEG Acquisition

EEG signals were collected with two systems: a **Neuroscan SynAmps** system (1000 Hz sampling, 0.3–15 Hz bandpass) and a **BrainProducts HD-EEG** system (400 Hz sampling, 0.1 Hz high-pass). To harmonize the datasets, 17 common electrodes were retained (C3, C4, Cz, F3, F4, F7, F8, Fp1, Fp2, Fpz, Fz, O1, O2, Oz, P3, P4, Pz). Post-awakening recordings were discarded; only pre-awakening segments were preserved.

### 2.3 EEG Preprocessing

EEG recordings in EDF format were resampled to 400 Hz for uniformity. Unwanted channels were removed, followed by z-score normalization per channel. The signals were converted into 2D arrays (17 × time length) and stored as NumPy arrays.

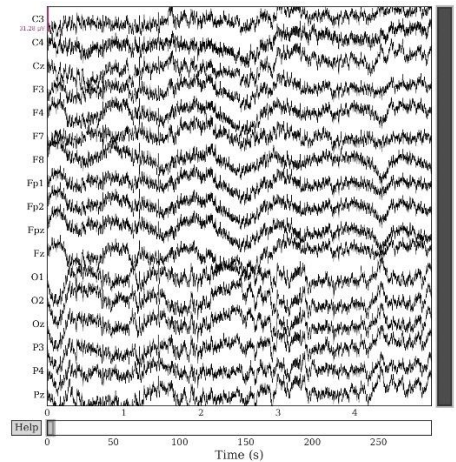

*Fig. 2 – Example EEG segment recorded during the 300 seconds preceding awakening from REM sleep. Data are shown from 17 harmonized electrodes (e.g., C3, C4, F3, O1, O2), resampled at 400 Hz and normalized by z-score per channel.*

Segments were extracted at different durations: T-15s (15 s before awakening), T-30s, T-60s, and T-total (up to 120 s before awakening). No additional filtering was applied to preserve the raw signal integrity.

We chose to retain these time segments before awakening because various studies correlate the occurrence of dreaming with the seconds preceding awakening. Twenty seconds before waking, a reduction in low-frequency waves (1–4 Hz) is recorded, whereas an increase in these low-frequency waves in the same region is associated with the absence of dreaming, whether during REM sleep or not (Siclari et al., 2017).

### 2.4 Dream Reports

Participants were awakened at different sleep stages (N1, N2, N3, REM). Each awakening produced an oral or written dream report, transcribed verbatim without modifications. Raw transcriptions exhibited lexical variability and inconsistencies due to the spontaneous nature of the reports. To address this, a condensed one-sentence description was also created, offering a clearer summary.

### 2.5 Image generation and selection

The image production pipeline consists of sequential steps, each managed by a specialized AI agent. The first step was semantic extraction, identifying key dream elements (emotional nuances, context, colors, individuals, internal and external relationships) to build a clear representation of essential components. The second step was prompt creation for the image generation model. These instructions included established aesthetics and defined characteristics (image format, photorealism level, representation of relationships, colors, people, etc.).

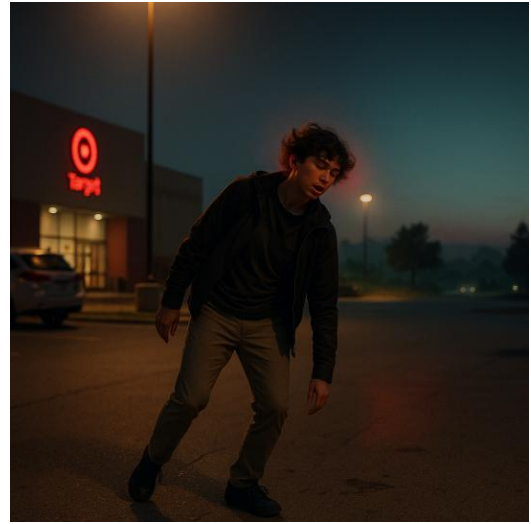

*Fig. 3 – AI-generated visual representation of a participant's dream. The image was produced using a multi-step pipeline including semantic extraction, prompt creation, neuropsychological validation, and iterative optimization with DALL·E 3.*

The third step involved neuropsychological validation or adjustment by a specialized agent, which included rule-based checks (e.g., consistency with affective valence, presence of main actors, preservation of dream context). This aimed to reduce interpretive bias while maintaining semantic fidelity.

At this stage, optional human monitoring was possible, allowing researchers to adapt the prompt as needed. However, excessive human intervention could reduce image generation homogeneity.

The fourth step was image generation using the DALL·E 3 model to produce valid visual representations. The fifth step was fidelity evaluation, comparing the generated image to the original dream transcript (not the prompt). A fidelity score (range 0–5) was computed based on semantic similarity between the dream transcript and the generated image, using both automated NLP embeddings (e.g., cosine similarity of BERT sentence embeddings) and manual evaluation. Only samples reaching a minimum threshold of 3 were retained.

Due to the scarcity of EEG data linked to dream transcripts, pairs judged to have low fidelity were not discarded but reintroduced into the next iteration. The sixth step was the optimization loop, enabling prompt regeneration until a minimal fidelity threshold was met.

## 3 Dataset Description

### 3.1 Global statistics

The dataset includes approximately 31.105 hours of recordings with 129 dream narratives presented in two formats: raw transcript and condensed description. Each transcript is paired with an image, resulting in 129 images. The lexicon and vocabulary vary considerably due to collection across multiple subjects. Because transcripts were captured verbatim, they often contained disfluencies, incomplete sentences, and lexical inconsistencies, which is inherent to spontaneous dream reporting. To address this, a secondary "description" column was provided, offering a concise and optimized one-sentence summary.

### 3.2 File Structure

The dataset is organized as follows:

- A "name" column containing the entry label.

- An "image" column with the 512px image representing a faithful visual of the participant's dream.

- An "eeg_15s" column containing the last 15 seconds of brain activity before awakening.

- An "eeg_30s" column containing the last 30 seconds of brain activity before awakening.

- An "eeg_60s" column containing the last 60 seconds of brain activity before awakening.

- An "eeg_total" column containing up to the last 120 seconds of brain activity before awakening (variable depending on available data).

- A "transcription" column with the verbatim text of the dream experience as reported by the participant.

- A "description" column providing a simplified, optimized summary of the transcript.

## 4 Use Cases and Interdisciplinary Perspectives

The Dream2Image dataset paves the way for numerous potential applications far beyond simple data compilation. Its strength lies in the unprecedented alignment of three complementary modalities: EEG signals, textual transcripts, and AI-generated images.

This multimodal combination creates bridges across neuroscience, psychology, artificial intelligence, and cognitive science.

### 4.1 Fundamental Neuroscience Research

The primary application of this dataset is the study of the neural correlates of dreaming. Dream2Image provides a foundation for observing and understanding correlations between neural patterns and subjective dream content. Such observations may help establish links between altered states of consciousness and wakefulness, thereby deepening the understanding of unconscious processes and sleep. For instance, one could analyze how REM sleep brain activity translates into narrative and visual dream experiences.

### 4.2 Psychology and Psychiatry

The study of dreams through imaging provides a new therapeutic foundation. Indeed, the meaning of the dream, produced through interpretation, must be consistent with the entirety of behavioral data, which highlights the complexity of interpretation compared to the simple rationalization of brain states (Parot, 2007).

Dream analysis may contribute to the study of emotional disorders, psychological resilience mechanisms, and the potential development of dream-based interventions. Dreams play an active role in the processing of emotional memories (Zhang et al., 2024). Their study could also provide a form of access to the unconscious, a better understanding of psychic conflicts, and the design of targeted therapeutic approaches (Ruby, 2011).

Particular attention should be paid to the deeper stages of sleep, as "the further one progresses in the sleep cycles, the more dreams incorporate distant memories, such as those from childhood" (Oudiette, 2025).

One other promising application concerns emotions and dream valence. Combining EEG data with dream reports enables the examination of neural signatures of dream emotions. This could lead to various clinical applications, such as studying recurrent nightmares, their origins, and neurophysiological impacts. Potential directions also include PTSD treatment strategies through the analysis and visual representation of dreams. Finally, with AI support, it may become feasible to develop non-invasive mental health monitoring tools via sleep.

### 4.3 AI & Machine Learning

The dataset was designed for AI models in visual and EEG decoding. While it offers strong potential for multimodal training, its relatively small size poses risks of overfitting and should be complemented with data augmentation or transfer learning. The main aim is to open new research avenues or improve existing algorithms through training or fine-tuning. The dataset is also suitable as a benchmark for inter-model comparisons in decoding and encoding tasks. It holds strong potential for training ML/DL models on multimodal decoding tasks linking EEG, text, and images.

### 4.4 Brain–Computer Interfaces (BCI)

By linking EEG signals to generated images, Dream2Image opens the possibility of interfaces capable of translating brain activity into visually comprehensible representations. Long-term applications include assisted communication for paralyzed patients and creative expression tools, enabling individuals to "draw with their brain."

## 5 Ethical Considerations

All data were collected under strict informed consent and anonymized. Generated images may reflect biases inherent to AI models and should therefore be interpreted with caution. Code and preprocessing scripts are provided to facilitate reproducibility. All data underwent rigorous anonymization to protect participant privacy. No dataset information allows identification of individuals. Names, labels, and other identifiers were removed during initial filtering and replaced with generalized terms during editing. All data were processed internally and not disseminated or used in commercial tools.

It is important to note that EEG signals exhibit high inter-subject variability and are subject to biases. These data provide a basis for hypotheses and interpretations but do not represent an exact reflection of the studied elements.

## 6 Mechanisms

The nature of the visual stimulus generated relies, on the one hand, on an active selection mechanism based on reactivation that prioritizes high-value associations (Oudiette et al., 2013). More specifically, this involves memory consolidation (Diekelmann & Born, 2010), hippocampal reactivation (Foster & Wilson, 2006), the sleep–learning relationship (Karni et al., 1994), and finally, memory selection mechanisms (Fischer & Born, 2009).

However, although memory reactivations and dream experiences share certain characteristics—such as the incorporation of fragments of recent experiences, a bias toward salient and novel memories, and some underlying neural correlates—significant disparities remain. In its classical definition, it is unlikely that rapid, spontaneous hippocampal-induced neural replay constitutes the primary neural basis of dreaming (Picard-Deland et al., 2023). Furthermore, some studies suggest that the dream experience may be characterized by distinct local EEG activity patterns within a posterior cortical "hot zone," regardless of the brain's overall state or the ability to recall dream content (Siclari et al., 2017).

The nature of the visual stimulus generated also depends on the quality of dream recall, which can be influenced by various factors such as the stage of sleep. Dreams are recalled more frequently during REM sleep than during N3 sleep, and slightly more often during REM than during N2 sleep (Picard-Deland et al., 2022).

Finally, the nature of the visual stimulus generated is also influenced by the intensity of the dream and its variations in form, such as lucid dreaming, which may contribute to better outcomes due to increased activity in the Anterior Prefrontal Cortex (aPFC), the parietal cortex, and the Frontoparietal Control Network (Baird et al., 2019).

## 7 Discussion & Conclusion

In this work, we sought to create a foundation for the study of dreams using EEG and AI-generated images. The aim was to address the scarcity of such data in the scientific literature and to introduce a new methodological approach. We found it is indeed feasible to construct such a dataset, but the main challenge remains the limited availability of EEG data recorded during sleep alongside dream transcripts. Collecting such data requires extensive facilities and considerable resources, yielding only small amounts of material.

Our results demonstrate the feasibility of generating visual representations of dream experiences, although their fidelity depends strongly on the quality of dream recall and the constraints of current generative models. The main limitation is the quality of dream recall, which can be significantly improved with preparation and training, combined with relaxation, breathing, or meditation exercises. These findings are of considerable importance for neuropsychology, as they clearly open a path toward studying dreams with AI applied to sleep analysis.

We also note that it is more relevant to work with segments from individuals in REM sleep, as these dreams are more elaborate and affective, more visually and kinesthetically engaging, more closely connected to waking life, and contain more visual imagery (Nielsen, 2000).

However, this dataset also presents important limitations due to its relatively small sample size compared to ML/DL requirements. This limitation increases the risk of overfitting and restricts the generalizability of trained models. Future work should explore larger-scale, collaborative datasets, the use of transfer learning, and data augmentation techniques to compensate for the current scarcity. Despite the limited number of samples available, this number appears sufficient for in-depth work, since the study of dreams with EEG must focus on more subtle features of brain activity across space and time (Nir & Tononi, 2010). We therefore recommend continued experimental protocols in sleep research that include systematic dream report collection, forming the basis for future dream studies. Collaborative projects such as this one offer great richness through population diversity and protocol variety, enabling greater generalizability and more promising results.

# 8  References

Baird, B., Mota-Rolim, S. A., & Dresler, M. (2019). The cognitive neuroscience of lucid dreaming. *Neuroscience & Biobehavioral Reviews, 100,* 305–323. https://doi.org/10.1016/j.neubiorev.2019.03.008

Diekelmann, S., & Born, J. (2010). The memory function of sleep. *Nature Reviews Neuroscience, 11*(2), 114–126. https://doi.org/10.1038/nrn2762

Fischer, S., & Born, J. (2009). Anticipated reward enhances offline learning during sleep. *Journal of Experimental Psychology: Learning, Memory, and Cognition, 35*(6), 1586–1593. https://doi.org/10.1037/a0017256

Foster, D. J., & Wilson, M. A. (2006). Reverse replay of behavioural sequences in hippocampal place cells during the awake state. *Nature, 440*(7084), 680–683. https://doi.org/10.1038/nature04587

Horikawa, T., Tamaki, M., Miyawaki, Y., & Kamitani, Y. (2013). Neural decoding of visual imagery during sleep. *Science, 340*(6132), 639–642. https://doi.org/10.1126/science.1234330

Karni, A., Tanne, D., Rubenstein, B. S., Askenasy, J. J. M., & Sagi, D. (1994). Dependence on REM sleep of overnight improvement of a perceptual skill. *Science, 265*(5172), 679–682. https://doi.org/10.1126/science.8036518

Konkoly, K. R., Appel, K., Chabani, E., Mangiaruga, A., Gott, J., et al. (2021). Real-time dialogue between experimenters and dreamers during REM sleep. *Current Biology, 31*(14), 3019–3025. https://doi.org/10.1016/j.cub.2021.01.026

Nielsen, T. A. (2000). A review of mentation in REM and NREM sleep: "Covert" REM sleep as a possible reconciliation of two opposing models. *Behavioral and Brain Sciences, 23*(6), 851–866. https://doi.org/10.1017/S0140525X0000399X

Nir, Y., & Tononi, G. (2010). Dreaming and the brain: From phenomenology to neurophysiology. *Trends in Cognitive Sciences, 14*(2), 88–100. https://doi.org/10.1016/j.tics.2009.12.001

Oudiette, D. (2025, May 15). Que sait-on vraiment de nos rêves ? *Polytechnique Insights*. https://www.polytechnique-insights.com/tribunes/neurosciences/que-sait-on-vraiment-de-nos-reves/

Oudiette, D., Antony, J. W., Creery, J. D., & Paller, K. A. (2013). The role of memory reactivation during wakefulness and sleep in determining which memories endure. *Journal of Neuroscience, 33*(15), 6672–6678. https://doi.org/10.1523/JNEUROSCI.5497-12.2013

Parot, F. (2007). De la neurophysiologie du sommeil paradoxal à la neurophysiologie du rêve. *Sociétés & Représentations, 23*(1), 195–212. https://doi.org/10.3917/sr.023.0195

Perogamvros, L., Baird, B., Seibold, M., Riedner, B., Boly, M., & Tononi, G. (2017). The phenomenal contents and neural correlates of spontaneous thoughts across wakefulness, NREM sleep, and REM sleep. *Journal of Cognitive Neuroscience, 29*(10), 1766–1777. https://doi.org/10.1162/jocn_a_01155

Picard-Deland, C., Bernardi, G., Genzel, L., Dresler, M., & Schoch, S. F. (2023). Memory reactivations during sleep: A neural basis of dream experiences? *Trends in Cognitive Sciences, 27*(6), 568–582. https://doi.org/10.1016/j.tics.2023.02.006

Picard-Deland, C., Konkoly, K., Raider, R., Paller, K. A., Nielsen, T., Pigeon, W. R., & Carr, M. (2022). The memory sources of dreams: Serial awakenings across sleep stages and time of night. *Sleep, 46*(4), zsac292. https://doi.org/10.1093/sleep/zsac292

Ruby, P. M. (2011). Experimental research on dreaming: State of the art and neuropsychoanalytic perspectives. *Frontiers in Psychology, 2,* 286. https://doi.org/10.3389/fpsyg.2011.00286

Siclari, F., Baird, B., Perogamvros, L., Bernardi, G., LaRocque, J. J., Riedner, B., Boly, M., Postle, B. R., & Tononi, G. (2017). The neural correlates of dreaming.

*Nature Neuroscience, 20*(6), 872–878. https://doi.org/10.1038/nn.4545

Wong, W., Andrillon, T., Decat, N., Herzog, R., Noreika, V., Valli, K., et al. (2023). *The DREAM database* [Dataset]. Monash University. https://doi.org/10.26180/22133105.v7

Zhang, J., & Wamsley, E. J. (2019). EEG predictors of dreaming outside of REM sleep. *Psychophysiology, 56*(7), e13368. https://doi.org/10.1111/psyp.13368

Zhang, J., Pena, A., Delano, N., Sattari, N., Shuster, A. E., Baker, F. C., Simon, K., & Mednick, S. C. (2024). Evidence of an active role of dreaming in emotional memory processing shows that we dream to forget. *Scientific Reports, 14*(1), 12845. https://doi.org/10.1038/s41598-024-58170-z